\documentclass[11pt,twocolumn]{article}
\usepackage{graphicx}
\usepackage{amssymb}
\usepackage{color}
\usepackage{titling}
\usepackage[]{epsfig}
\usepackage{bm}
\usepackage{hyperref}
\begin{document}

\title{\vspace{-3.0cm}New Hadrons Discovered at BESIII}
\author{Zhiqing Liu$^a$, Ryan~E.~Mitchell$^b$ \\
$^a$Shandong University, Jinan 250100, China \\
$^b$Indiana University, Bloomington 47401, USA
}
\date{\today}
\maketitle

\begin{figure*}[htb]
\begin{center}
\includegraphics*[width= 2.0\columnwidth]{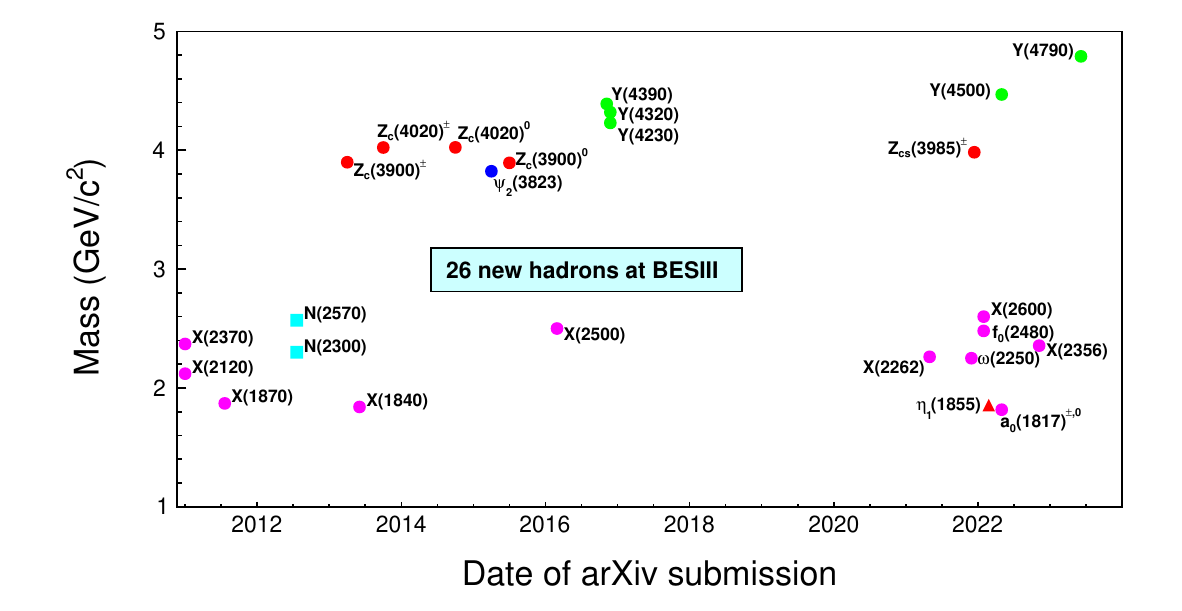}
\caption{\label{fig:hadrons}(Color online) The 26 new hadrons discovered by the BESIII experiment, sorted by their mass and year of discovery.  
Red circles indicate states with exotic flavor combinations decaying into heavy mesons; 
the blue circle is a state consistent with a conventional $c\bar{c}$ meson; 
the green circles are states produced directly in $e^+e^-$ collisions with no clear interpretation; 
the magenta circles are new light states decaying into mesons; 
the light blue circles are new light baryon states;
and the red triangle is a state with exotic $J^{PC}$.
Data is quoted from the BESIII Collaboration: \url{http://english.ihep.cas.cn/bes/re/pu/NewParticles/}.
}
\end{center}
\end{figure*}

The Beijing Spectrometer~(BESIII) experiment, which has been studying particle collisions produced by the Beijing Electron Positron Collider~(BEPCII) since 2009, has so far discovered a total of 26 new particle candidates (Fig.~\ref{fig:hadrons}).  
The particles, all strongly-interacting hadrons, are thought to be composed of quarks~($q$), antiquarks~($\bar{q}$), and gluons~($g$) bound together by the strong force in a surprising variety of configurations.  
Elucidating their properties provides new insight into the complex nature of nuclear interactions and has been one of the central goals of the BESIII collaboration, which consists of an international team of more than 600 scientists.  
While several of the newly discovered hadrons are consistent with the conventional two-quark picture of mesons~($q\bar{q}$) or the three-quark picture of baryons~($qqq$), and were expected, others are more surprising and have exotic configurations, 
such as tetraquarks~($qq\bar{q}\bar{q}$), meson molecules~($(q\bar{q})(q\bar{q})$), glueballs~($gg$), or hybrid mesons~($qg\bar{q}$).
They have led to new advances in our theoretical understanding of the strong force,
for example through the development of new approaches in quantum field theory~\cite{Mai:2022eur} or understanding the importance of hadron molecules~\cite{Guo:2017jvc},
and have initiated intense experimental effort~\cite{Chen:2022asf}.
In the following, we briefly tour the 26 new hadrons, roughly from heaviest to lightest.  Their names appear in boldface the first time they are mentioned.

The heaviest particles produced at BEPCII, with masses above 3~GeV/$c^2$ (about three times the mass of a proton), contain at least one charm quark~($c$) and one charm antiquark~($\bar{c}$) and are referred to as charmonium.  The easiest charmonium states to produce at BEPCII share the same spin~($J$), parity~($P$), and symmetry under charge conjugation~($C$) as the photon~($J^{PC}=1^{--}$). They can thus be produced directly in reactions such as $e^+e^-\to \gamma^* \to J/\psi$~(the $J/\psi$ being the first charmonium state ever discovered, in 1974) when the $e^+e^-$ center-of-mass energy matches the mass of the particle being produced.
While the $J/\psi$ is simply a $c\bar{c}$ pair with quark spins arranged to have one unit of angular momentum, 
at higher $e^+e^-$ energies more exotic particles apparently emerge.  
The reaction $e^+e^- \to \pi^+\pi^- J/\psi$, for example, shows two enhancements in the production rate as the $e^+e^-$ energy is scanned between 3.77 and 4.60~GeV.  
One enhancement is at 4.23~GeV, named the ${\bm{Y(4230)}}$, and the other is at 4.32~GeV, named the ${\bm{Y(4320)}}$~\cite{Y4230Y4320}.  They cannot be described as conventional $c\bar{c}$ charmonium states and their internal structure remains unclear.
Other particle candidates have been found in a similar fashion, and are also sources of intense speculation: 
the ${\bm{Y(4390)}}$~\cite{Y4390} was discovered in 
$e^+e^- \to Y(4390) \to \pi^+\pi^- h_c$;
the ${\bm{Y(4500)}}$~\cite{Y4500} in
$e^+e^- \to Y(4500) \to K^+K^- J/\psi$;
and at even higher energies
the ${\bm{Y(4790)}}$~\cite{Y4790} in
$e^+e^- \to Y(4790) \to D_s^{*+}D_s^{*-}$.

Another class of particles is even more surprising; these particles appear as decay products in $Y$ decays and can carry electric charge.  
The first in this class discovered at BESIII was the 
${\bm{Z_c(3900)^\pm}}$~\cite{ZC3900}, which appears in the multi-step reaction
$e^+e^- \to Y(4230) \to Z_c(3900)^\pm \pi^\mp \to \pi^+ \pi^- J/\psi$.  Because the $Z_c(3900)^\pm$ decays to a $J/\psi$
and has a mass in the charmonium region,
it must contain a $c\bar{c}$ pair.  But since the $Z_c(3900)^\pm$ has an electric charge, and since the $c\bar{c}$ pair is electrically neutral, more quarks must be involved.
The most accepted interpretation is that the $Z_c(3900)^+$ contains four quarks: 
$c\bar{c}u\bar{d}$,
although the configuration of the up~($u$) and down~($d$) quarks relative to the charm quarks is still under investigation.
Others in this class include 
the  ${\bm{Z_c(4020)^\pm}}$~\cite{ZC4020a,ZC4020b}
and the ${\bm{Z_{cs}(3985)^\pm}}$~\cite{ZCS3985}, which also includes a strange~($s$) quark.
Electrically neutral partners to these states, including the 
${\bm{Z_c(3900)^0}}$~\cite{ZC3900n} and
${\bm{Z_c(4020)^0}}$~\cite{ZC4020n}, have also been discovered at BESIII and lend further support to their four-quark interpretation.
One property shared by all of the $Z$ states discovered at BESIII is that 
their masses are just a bit higher than that required to 
decay to a pair of mesons each containing one charm quark.
For example, the mass of the $Z_c(3900)$ is just enough so that it can decay to a $D$ and a $D^*$ meson.
This suggests they may be meson molecules with configurations such as $(c\bar{d})(\bar{c}u)$, where the parentheses indicate more closely bound $q\bar{q}$ combinations.

Unlike the discoveries of the $Y$, $Z_c$, and $Z_{cs}$ states, 
the discovery of the ${\bm{\psi_2(3823)}}$~\cite{PSI23823}
aligns with expectations from a simple hydrogen-like model of the $c\bar{c}$ system.  
In this case the quark spins contribute one unit of spin angluar momentum, the quarks rotate with two units of orbital angular momentum, and the whole system has two units of total angular momentum. 
The $\psi_2(3823)$ therefore serves as an invaluable candle with which other particles can be compared.

In the light hadron sector, 
where the charm quarks in charmonium are replaced by the lighter up~($u$), down~($d$), and strange~($s$) quarks, 
radiative decays of the $J/\psi$ meson~($J/\psi \to \gamma gg \to \gamma + \mathrm{hadrons}$) 
provide an environment rich in gluons and ripe for discovery.
BESIII has recently collected the world's largest sample of $J/\psi$ decays~(10~billion),
and is therefore in a position to perform cutting edge searches for exotic light hadrons like hybrid
mesons, glueballs, and hadron molecules.
The decay $J/\psi \to \gamma \pi^+\pi^-\eta^\prime$ has proved especially fruitful.
BESIII has not only confirmed the BES discovery of the $X(1835)$ near $p\bar{p}$ threshold in this process,
but has also discovered the heavier ${\bm{X(2120)}}$~\cite{X2120X2370}, ${\bm{X(2370)}}$~\cite{X2120X2370}, 
and ${\bm{X(2600)}}$~\cite{X2600}, all decaying to $\pi^+\pi^-\eta^\prime$.
In related processes, BESIII has discovered other structures near $p\bar{p}$ threshold, including the 
${\bm{X(1840)}}$~\cite{X1840} in $J/\psi\to\gamma X(1840)\to \gamma (6\pi)$ and the
${\bm{X(1870)}}$~\cite{X1870} in $J/\psi\to\omega X(1870)\to \omega (\pi^+\pi^-\eta)$.

Glueball states are expected to have masses between 1.5 and 2.6~GeV/$c^2$ and are expected to be 
prominent in decays to two identical hadrons.
In this regard, the BESIII discoveries of the 
${\bm{f_0(2480)}}$~\cite{F02480} in 
$J/\psi \to \gamma f_0(2480) \to \gamma (\eta^\prime \eta^\prime)$
and the ${\bm{X(2500)}}$~\cite{X2500} in 
$J/\psi\to \gamma X(2500) \to \gamma (\phi \phi)$
provide important markers in the glueball spectrum.
In addition to decays of the $J/\psi$, complementary information about the glueball spectrum comes from other experimental processes.
A new $a_0$-like state, seen as both ${\bm{a_0(1817)^0}}$ and ${\bm{a_0(1817)^\pm}}$~\cite{A01817}, was discovered in $D$ meson decays. 
If the $a_0(1817)$ is an isospin-one partner of the $f_0(1710)$, as appears likely, this would rule out a glueball interpretation for the 
$f_0(1710)$.

While light hadrons with quantum numbers $J^{PC}$ allowed by conventional $q\bar{q}$ mesons
are difficult to classify as either conventional~(i.e. $q\bar{q}$) or not~(i.e. non-$q\bar{q}$),
hadrons with exotic quantum numbers must be non-$q\bar{q}$.
This is the case for the ${\bm{\eta_1(1855)}}$~\cite{ETA11855} discovered in the process $J/\psi \to \gamma \eta_1(1855) \to \gamma (\eta\eta^\prime)$.
The exotic $J^{PC}=1^{-+}$ quantum numbers of the $\eta_1(1855)$ make it a strong candidate
for a hybrid meson or hadron molecule.

New excited light meson states have also been found in detailed $e^+e^-$ energy scans.
A peaking structure was observed in a scan of the process $e^+e^- \to \omega \pi^0\pi^0$
at an $e^+e^-$ center-of-mass energy of around 2.25~GeV. 
This structure, called ${\bm{\omega(2250)}}$~\cite{OMEGA2250},
is possibly an excitation of the ground-state $\omega$ meson.
Other new particle candidates, the ${\bm{X(2262)}}$~\cite{X2262} and the ${\bm{X(2356)}}$~\cite{X2356}, 
were discovered in the $\Lambda\bar{\Lambda}$ system using an energy scan above 3~GeV.
These two candidates are produced along with a $\phi$ meson and an $\eta$ meson, respectively,
thus they must have different $C$-parity quantum numbers.

The spectrum of light baryon states can be accessed at BESIII through the world's largest
sample of $\psi(3686)$ decays~(2.7~billion).  A study of the process 
$\psi(3686) \rightarrow p\bar{p}\pi^0$, for example,
revealed the existence of two new baryons, 
the ${\bm{N(2300)}}$ and ${\bm{N(2570)}}$~\cite{N2570N2300}.
Many more processes, which are sensitive to a rich variety of previously unseen baryon states,
are currently under study.

Starting from the summer of 2024, the BEPCII accelerator will be upgraded to BEPCII-U, expanding the center-of-mass energy reach to 5.6~GeV
and providing up to a factor of three improvement in intensity.
Both upgrades will 
allow more detailed studies of the 26 particles so far discovered~(for example measuring particle $J^{PC}$, determining properties of scattering amplitudes, and so on)
and will
expand the discovery potential of BESIII into new territories.
In addition to discovering new particles, the BESIII experiment has a physics program that spans a wide variety of topics ranging from strong force dynamics to precision measurements of the electroweak force to searches for physics beyond the Standard Model of Particle Physics.  The collaboration has now published more than 500 papers.

\end{document}